\documentclass{moriond}

\def\be{\begin{equation}}
\def\ee{\end{equation}}
\def\bea{\begin{eqnarray}}
\def\eea{\end{eqnarray}}

\newcommand{\Photo}{}

\begin{document}
\begin{flushright}DO-TH 21/15\end{flushright}
\vspace*{4cm}
\title{SYNERGIES OF TOP AND \boldmath$B$ ANOMALIES IN SMEFT}

\author{Stefan~Bi{\ss}mann \footnote{Speaker}, Cornelius~Grunwald, Gudrun~Hiller, Kevin~Kr{\"o}ninger}
%\author{Cornelius~Grunwald}
%\author{Gudrun~Hiller}
%\author{Kevin~Kr{\"o}ninger}
\address{Fakult\"at Physik, TU Dortmund, Otto-Hahn-Str.4, D-44221 Dortmund, Germany}

\maketitle\abstracts{
We present constraints on Wilson coefficients from a combined fit to data from top-quark and beauty-physics measurements within the Standard Model effective field theory. 
We consider present data on top-quark pair production and decay, $b\to s$ flavor changing neutral currents, and $Z\to b\bar b$ decay, as well as different future scenarios.
These comprise projections for HL-LHC, Belle II, and a lepton collider using the example of CLIC.
We investigate opportunities for detecting deviations from the Standard Model hinted at by present data on $b\to s$ transitions.
In the fit, we find strong synergies between top-quark and beauty physics that allow to tighten constraints on various coefficients.
}

\section{Effective theories for new physics}

Due to the absence of any direct observation of physics beyond the Standard Model (BSM), present data suggest that BSM particles could very well be too heavy to be directly detected by experiments at the LHC. 
This opens up a different approach well known from flavor physics and recently applied to top-quark physics: indirect searches employing effective field theories (EFT).
In this regard, the Standard Model EFT (SMEFT) has become very popular as it allows model-independent searches for BSM physics combining different physics sectors in one consistent framework.
The key concept of this framework is to add higher-dimensional operators $O_i$ consistent with the Standard Model (SM) symmetry groups to the SM Lagrangian. 
Each of the operators appears together with a corresponding Wilson coefficients $C_i$ and is suppressed by inverse powers of the scale of BSM physics $\Lambda$.
Leading contributions to LHC physics obeying lepton and baryon number conservation arise at dimension six:
\begin{equation}
       \mathcal{L}_\textmd{SMEFT}=\mathcal{L}_\textmd{SM}+\sum_i\frac{C^{(6)}_i}{\Lambda^2}O_i^{(6)}+\mathcal{O}\left(\Lambda^{-4}\right)\,.
    \label{eq:L_SMEFT}
\end{equation}
In this framework, deviations from the SM correspond to non-zero Wilson coefficients.

$SU(2)_L$ invariance of the SMEFT Lagrangian links up-type and down-type quarks. 
Rotating quark fields from the flavor basis $q^i_{L/R}$ to the mass basis $q_{L/R}^{\prime i} = \left(S^{q^\dagger}_{L/R}\right)_{ij} q^{ j}_{L/R}$ connects different physics sectors with $q=u,d$ and flavor indices $i,j=1,2,3$.
The CKM matrix reads
\begin{equation}
    V = \left(S^u_L\right)^\dagger S^d_L\,.    
\end{equation}
While rotations of right-handed quark fields can simply be absorbed in the definition of the Wilson coefficients, contributions involving quark doublets relate up and down sector via CKM-matrix elements.
Consider the following example:
\begin{equation}
    C^{ij}_{eq} O^{ij}_{eq}
    = \hat C^{kl}_{eq} \left(\bar e_R\gamma_\mu e_R\right) \left(  \bar u_L^{\prime k} \gamma^\mu  u_L^{\prime l} + V^\dagger_{mk} V^{\vphantom{\dagger}}_{ln} \bar d_L^{\prime m} \gamma^\mu  d_L^{\prime n}\right)\,,
    \label{eq:mass-operator-relation}
\end{equation}
where $\hat C_{e q}^{(1)kl} = C^{ij}_{eq} \left(S^{u\vphantom{d}}_L\right)^\dagger_{ki} \left(S^{u\vphantom{d}}_L\right)^{\vphantom{\dagger}}_{jl}$ denotes Wilson coefficients in the mass basis.
Here, we work in the up-type basis where CKM matrix elements appear in interactions of down-type quarks.

In the following, we apply this framework to a combined analysis of present top-quark, $Zb\bar b$, and $B$ data. 
In addition, we consider future projections for measurements at HL-LHC, Belle II, and a future lepton collider using the example of CLIC.
We only consider contributions from operators with third generation quarks (in the flavor basis), i.e. only $\hat C_i^{33}\neq0$, that interfere with the SM process.
We use the convention $\tilde C_i = \hat C_i^{33} v^2/\Lambda^2$ with $v=246\,\textmd{GeV}$ being the Higgs vacuum expectation value. 
In total, we include the following eleven coefficients in our analysis:
\begin{equation}
		\tilde C_{uB}\,,\enskip
		\tilde C_{uG}\,,\enskip
		\tilde C_{uW}\,,\enskip 
		\tilde C^{(1)}_{\varphi q}\,,\enskip
		\tilde C^{(3)}_{\varphi q}\,,\enskip
		\tilde C_{\varphi u}\,,\enskip
		\tilde C_{eu}\,,\enskip
        \tilde C_{lu}\,,\enskip
        \tilde C_{qe}\,,\enskip
        \tilde C^{(1)}_{lq}\,,\enskip 
		\tilde C^{(3)}_{lq}\,,
		\label{eq:future-SMEFT-coff}
\end{equation}
where we assume lepton flavor universality. 
While this is not a strong assumption presently as most $b\to s \ell^+\ell^-$ data is muon-specific, it is important once $b\to s\nu\bar\nu$ modes are included and those from $e^+e^-$ colliders.
To combine top-quark and $B$-physics data \cite{Bissmann:2019gfc}, we employ the \texttt{wilson} package \cite{Aebischer:2018bkb} to match SMEFT onto the low-energy weak effective theory \cite{Dekens:2019ept} (WET) and to include renormalization group evolution \cite{Alonso:2013hga}. 
SMEFT coefficients are given at the scale $\mu=1\,\textmd{TeV}$.

\section{Fit to present data}
We consider measurements of top-quark physics, namely total cross sections of pair production processes at the LHC ($t\bar t$, $t\bar t\gamma$, and $t\bar t Z$), the total decay width of the top quark, and $W$-boson helicity fractions. 
We further include data on $Z \to b \bar b$ transitions and $B$ physics in the form of total and differential branching ratios, angular distributions, and asymmetries \cite{Bissmann:2020mfi}.
Angular distributions on $b\to s \mu^+\mu^-$ data are of particular interest as they hint at deviations from the SM.
In total, this allows us to constrain the following eight SMEFT coefficients:
\begin{equation}
    \tilde C_{uB}\,,\quad \tilde C_{uG}\,,\quad \tilde C_{uW}\,,\quad \tilde C_{\varphi q}^{(1)}\,,\quad \tilde C_{\varphi q}^{(3)}\,,\quad  \tilde C_{\varphi u}\,,\quad \tilde C_{qe}\,,\quad \tilde C_{lq}^+\,,
    \label{eq:max-dof}
\end{equation}
where we also introduced the useful linear combinations $\tilde C^\pm_i = \tilde C^{(1)}_i \pm \tilde C^{(3)}_i$ to avoid flat directions in the multidimensional parameter space.
\begin{figure}[t]
    \centering
    \includegraphics[width=0.45\textwidth]{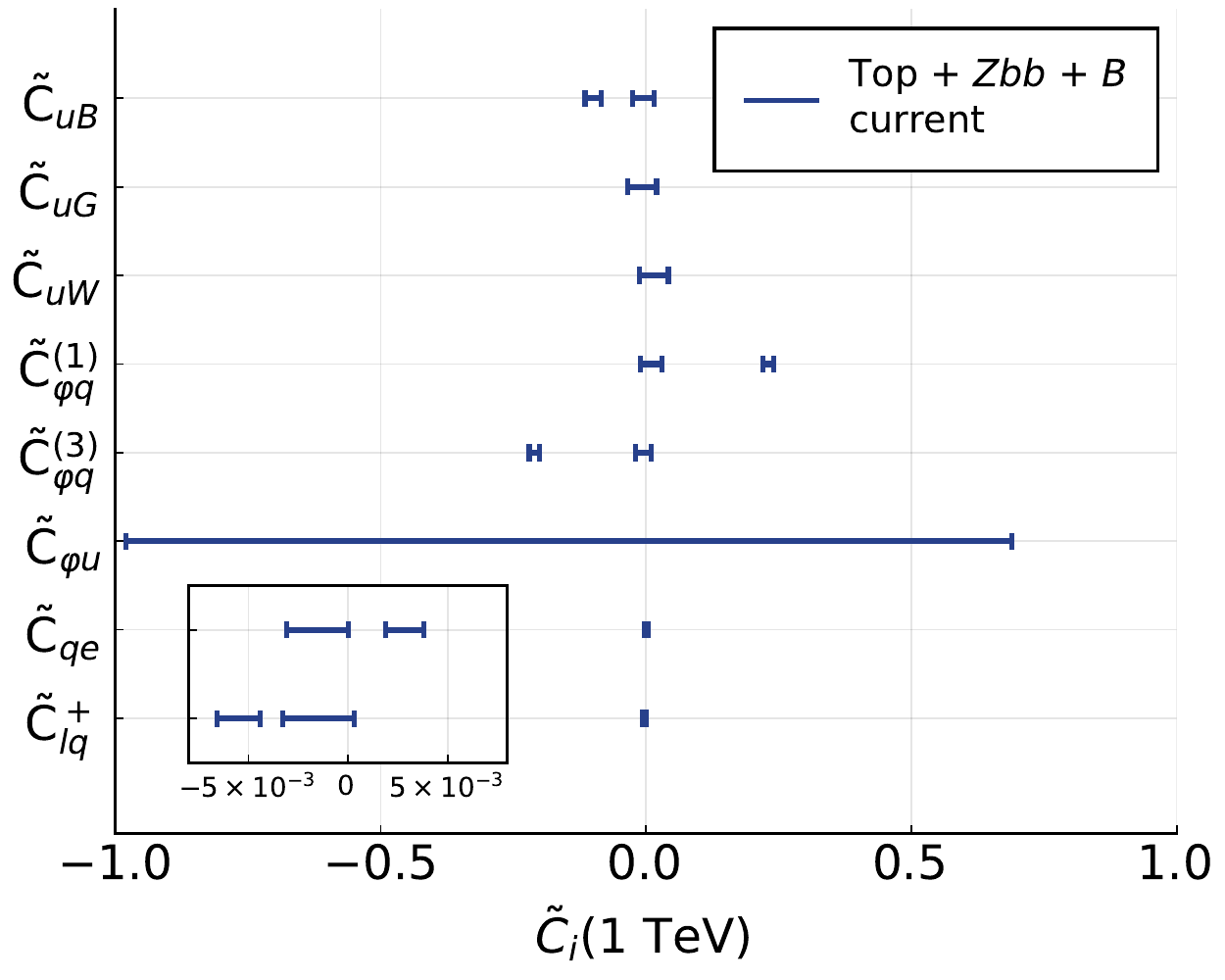}
    \includegraphics[width=0.45\textwidth]{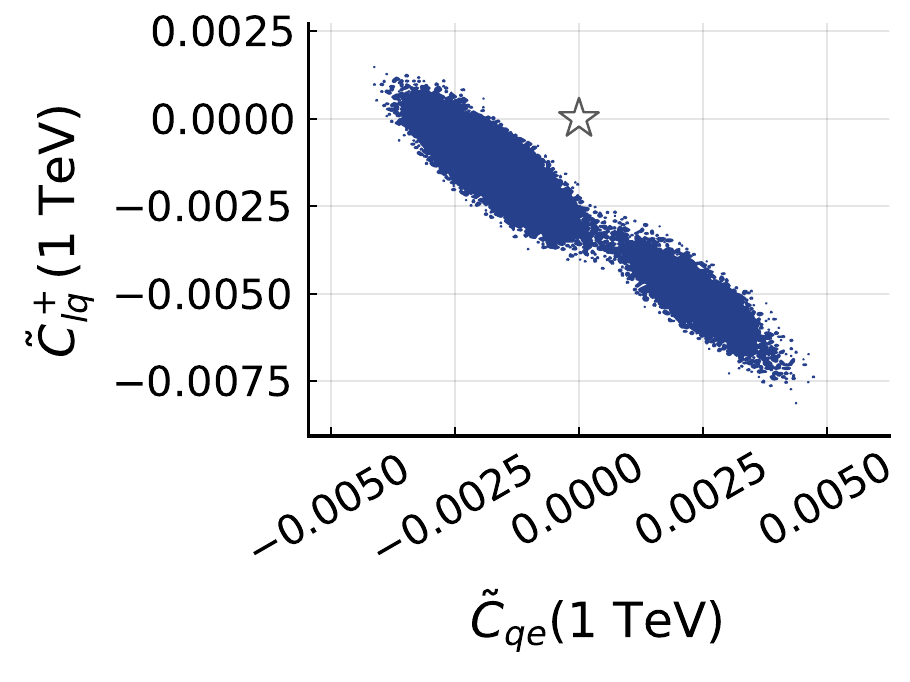}
    \caption{Constraints on SMEFT coefficients $\tilde C_i$ in Eq.~(\ref{eq:max-dof}) from a fit of eight coefficients to top-quark data, $Zbb$ data, and $B$-physics data. Shown are the smallest intervals containing $90\,\%$ posterior probability (left) and the two-dimensional projection of the posterior distribution in the $\tilde C_{qe}$-$\tilde C^+_{lq}$ plane (right). 
    The star denotes the SM. Plots from Ref.~\protect\cite{Bissmann:2020mfi} or adopted.
    } 
    \label{fig:TB-Now-maxDof}
\end{figure}

To derive constraints on Wilson coefficients in a Bayesian reasoning, we employ the EFT\textit{fitter} tool \cite{Castro:2016jjv}. 
For all coefficients we consider a uniform distribution over the interval $-1\leq \tilde C_i \leq 1$ as the prior distribution.
In Fig.~\ref{fig:TB-Now-maxDof} we show results from a fit of eight coefficients in Eq.~(\ref{eq:max-dof}) to data from top-quark measurements, $Zbb$ data, and $B$-physics.
We show both the one-dimensional projection (left) as well as the two-dimensional projection in the $\tilde C_{qe}$-$\tilde C^+_{lq}$ plane of the smallest 90\,\% region of the posterior distribution.
As can be seen in the one-dimensional projections, the strongest constraints are obtained for the four-fermion coefficients $\tilde C_{qe}$ and $\tilde C^+_{lq}$, $\mathcal{O}(10^{-3})$.
Constraints on $\tilde C_{uB}$, $\tilde C_{uG}$, $\tilde C_{uW}$, $\tilde C^{(1)}_{\varphi q}$, and $\tilde C^{(3)}_{\varphi q}$ are roughly one order of magnitude weaker with a total width of the smallest interval around $(5-7)\times 10^{-2}$. 
In contrast, $\tilde C_{\varphi u}$ remains almost unconstrained as constraints from $t\bar tZ$ data are weak due to limited precision and contributions to $B$ physics are strongly suppressed by the matching conditions. 
Interestingly, we find two solutions for several coefficients, which stem from correlations induced by matching the SMEFT basis onto the WET basis.
Inclusion of top-quark data allows to resolve these correlations \cite{Bissmann:2019gfc} to a certain degree.
However, in this case the sensitivity of top-quark data does not suffice to completely remove the non-SM branches from the posterior distribution.

The two semileptonic four-fermion coefficients are of particular interest as they are related to anomalies seen in $b\to s\mu^+\mu^-$ transitions.
As shown in the two-dimensional projection of the posterior distribution in the $\tilde C_{qe}$-$\tilde C^+_{lq}$ plane (Fig.~\ref{fig:TB-Now-maxDof}, right), we see deviations from the SM in these two coefficients.

\section{Future projections}
To estimate the impact of future experiments on constraints on SMEFT coefficients, we adopt estimates of the expected precision at HL-LHC \cite{Atlas:2019qfx}, Belle II \cite{Kou:2018nap}, and a future lepton collider at the concrete example of CLIC \cite{Abramowicz:2018rjq}. 
%To do so, we rescale uncertainties of present data according to the analyses of the experimental collaborations.
%When no explicit estimates are provided, we rescale statistical uncertainties according to the integrated luminosity and assume systematic uncertainties to be reduced by a factor of 2. 
%Similarly, we assume theory uncertainties to be also reduced by a factor of 2.
If no present measurement for an observable is available, we consider the SM value, and take into account central values of present data otherwise.

We distinguish between two scenarios: In the 'near future' scenario, we take into account projections for measurements at HL-LHC and Belle II. 
In this case, Belle II is of particular interest, as it is expected to give further insights in the $B$ anomalies.
%While these two experiments are already operating (Belle II) or planned to operate in the near future (HL-LHC), a lepton collider at the TeV scale has yet to be build. 
A lepton collider such as CLIC has yet to be decided to built, and thus we consider data from such an collider as a 'far future' scenario.

\begin{figure}[t]
    \centering
    \includegraphics[width=0.450\textwidth]{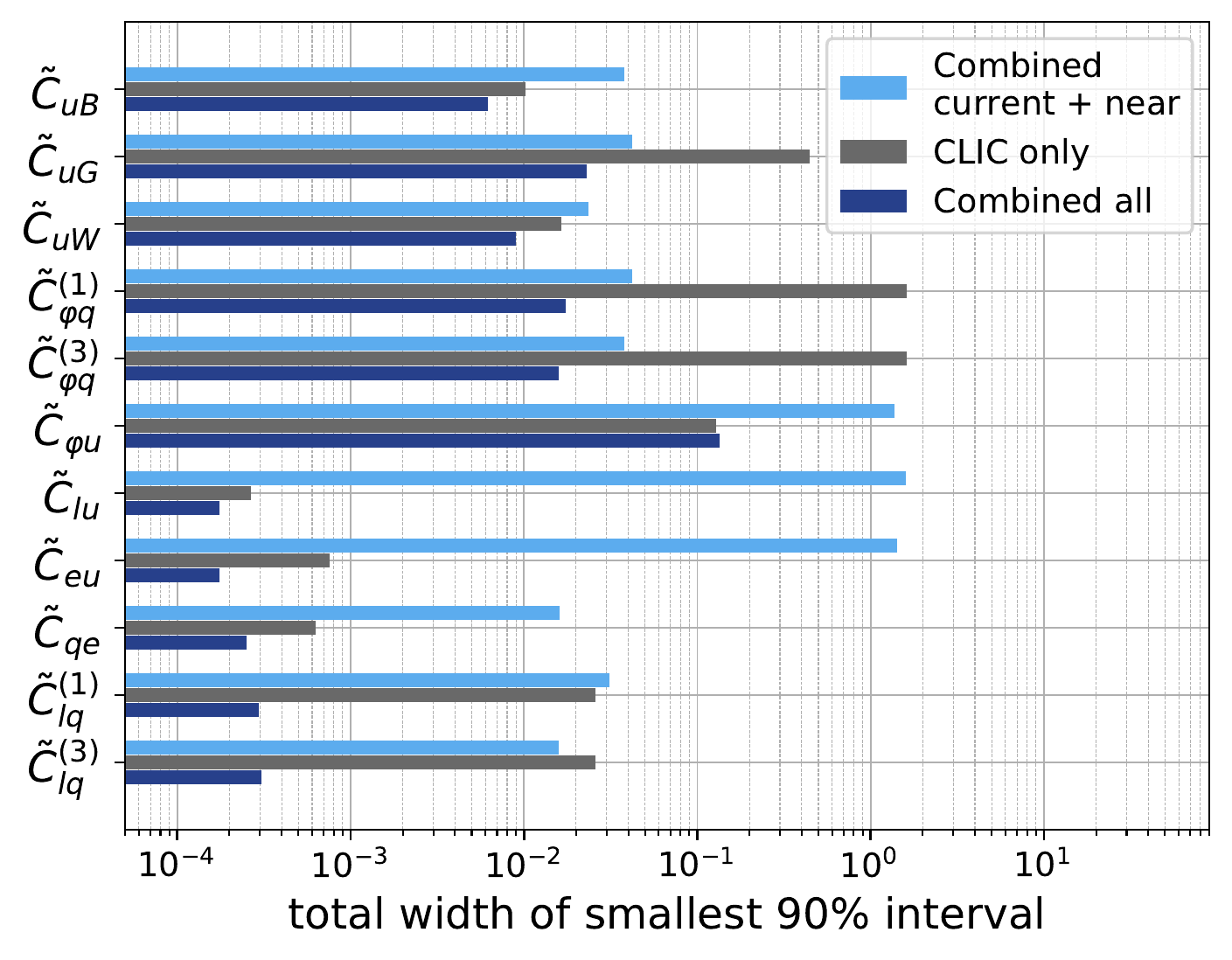}
    \includegraphics[width=0.45\textwidth]{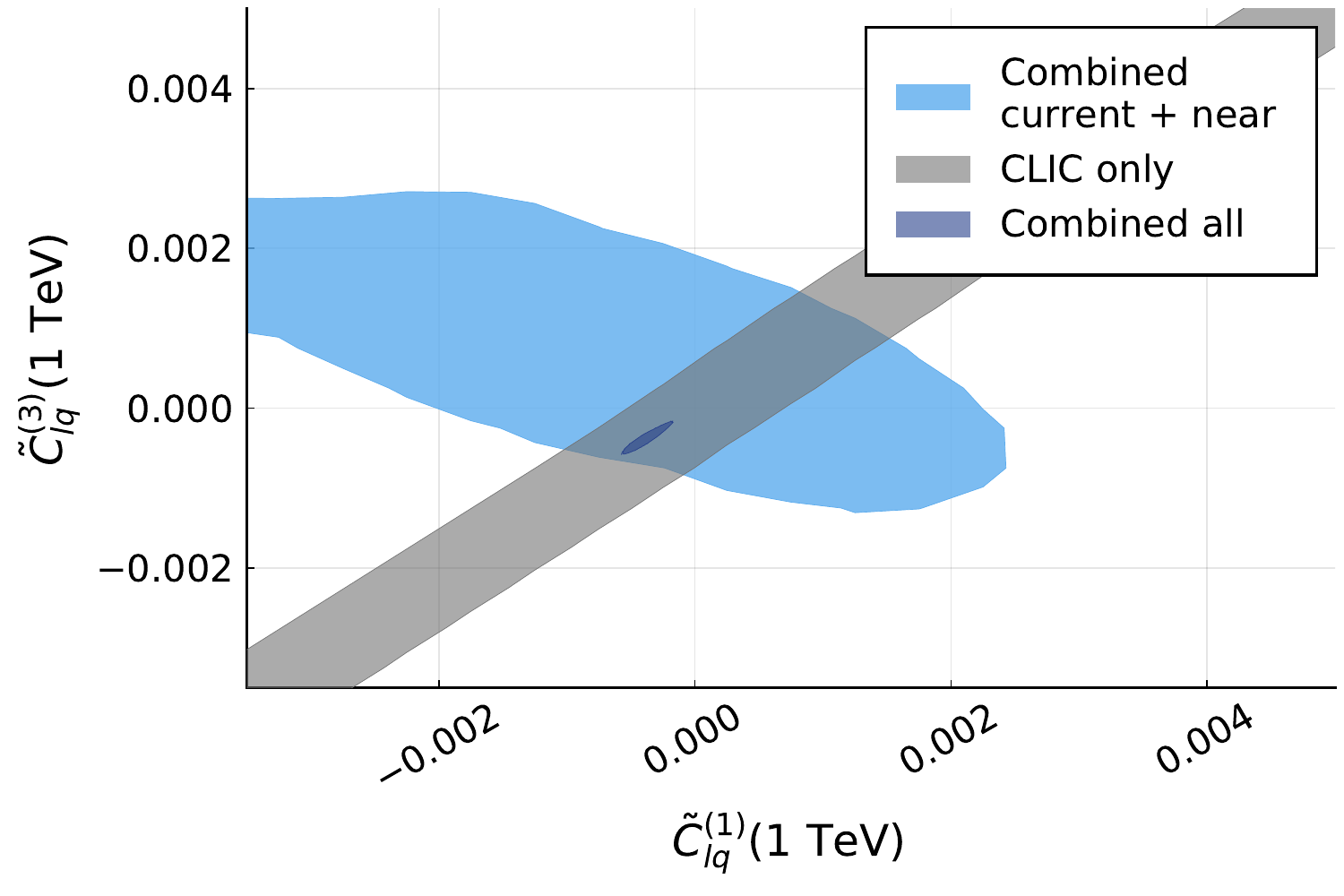}
    \caption{Constraints on SMEFT coefficients $\tilde C_i$ in Eq.~(\ref{eq:future-SMEFT-coff}) from a fit of all eleven coefficients to present data and projections. 
    Shown are the width of the smallest intervals containing $90\,\%$ posterior probability (left) and two-dimensional projections of the posterior distribution (right) from fits to 
    present data, HL-LHC and Belle II projections (light blue), CLIC projections (grey), and the combined set (blue). 
    Plots from Ref.~\protect\cite{Bissmann:2020mfi}.
    } 
    \label{fig:TB-far-maxDof}
\end{figure}

Combining observables from these experiments allows to test the complete eleven-dimensional parameter space of Wilson coefficients.
We show the results from a fit of all eleven coefficients to present data and near-future projections (light blue), CLIC projections (grey), and the combined set in Fig.~\ref{fig:TB-far-maxDof}.
As shown in the plot where the total width of the smallest 90\,\% interval are shown, the four-fermion coefficients can be constrained to a level of $\mathcal{O}(10^{-4})$.
In contrast, the weakest constraints are found for $\tilde C_{\varphi u}$, for which the width of the interval is around $10^{-1}$. 
Constraints on the remaining coefficients are at the level of $\mathcal{O}(10^{-2})$.
Combining present data with projections for HL-LHC, Belle II (especially $b\to s\nu\bar\nu$), and CLIC significantly improves the constraints on the SMEFT coefficients.
While in most cases one dataset is more sensitive to a specific operator and thus drives the constraints, we find an particularly significant enhancement by more than two orders of magnitude for both $\tilde{C}^{(1)}_{lq}$ and $\tilde{C}^{(3)}_{lq}$.
The reason is that the different sets of observables have different sensitivities: Constraints from CLIC are almost orthogonal to those obtained from present data and near-future projections for HL-LHC and Belle II.

This is shown in Fig.~\ref{fig:TB-far-maxDof} (right) where we give the two-dimensional projection of the posterior distribution in the $\tilde C^{(1)}_{lq}-\tilde C^{(3)}_{lq}$ plane obtained from fits to all eleven coefficients.
We find that both $\tilde{C}^{(1)}_{lq}$ and $\tilde{C}^{(3)}_{lq}$ show significant deviations from the SM in the combined fit assuming that Belle II data confirms present LHCb central values.

\section{Conclusions}
We presented results from fits of SMEFT coefficients to current top-quark, $Z\to b\bar b$, and $B$-physics data obtained in a fit of eight coefficients in Eq.~(\ref{eq:max-dof}) to the combined set. 
We discussed how the combination of the different datasets allows to tighten constraints on the SMEFT coefficients considered here. 
Exploiting the $SU(2)_L$-symmetry link between top-quark and beauty physics enabled us to constrain four-fermion operators involving top-quarks and charged leptons to a level of $\mathcal{O}(10^{-3})$. 
Due to hints for BSM physics in $b\to s \mu^+\mu^-$ transitions, we find deviations from the SM in two semileptonic four-fermion coefficients tested with present data, see Fig.~\ref{fig:TB-Now-maxDof}.

In the future, measurements at HL-LHC and Belle II will improve the constraints on several coefficients, while inclusion of top-quark data from a future lepton collider is paramount to remove flat directions in the parameter space and to constrain all eleven coefficients in Eq.~(\ref{eq:future-SMEFT-coff}) simultaneously, see Fig.~\ref{fig:TB-far-maxDof}.
Four-fermion coefficients are constrained at a level of $\mathcal{O}(10^{-4})$, and different sensitivities of top-quark and beauty data allow to see significant deviations from the SM in two coefficients assuming present central values on $b\to s\mu^+\mu^-$ data.

\section*{Acknowledgments}

SB is very grateful to the organizers to be given the opportunity to present this work.
CG is supported by the doctoral scholarship program of the \textit{Studienstiftung des deutschen Volkes}.

\end{document}